\documentclass[english,twocolumn]{revtex4}
\usepackage[T1]{fontenc}
\usepackage[latin1]{inputenc}
\usepackage{graphicx}

\makeatletter

\providecommand{\tabularnewline}{\\} 
\makeatletter
 
\usepackage{times}   
\makeatother 

\usepackage{babel} 
\begin{document}

\title{Calorimetry of Bose-Einstein condensates}

\author{P. B. Blakie$^{1}$, E. Toth$^{1}$, M. J. Davis$^{2}$ }

\address{$^{1}$Jack Dodd Centre for Photonics and Ultra-Cold Atoms, Department of Physics, University of Otago, Dunedin,
New Zealand\\
 $^{2}$ARC Centre of Excellence for Quantum-Atom Optics, School of
Physical Sciences, University of Queensland, Brisbane, QLD 4072, Australia}

\begin{abstract}
We outline a practical scheme for measuring the thermodynamic properties
of a Bose-Einstein condensate as a function of internal energy. We
propose using Bragg scattering and controlled trap manipulations to impart a precise amount of energy
to a near zero temperature condensate. After thermalisation the temperature
can be measured using standard techniques to determine the state equation
$T(U,N,\omega)$. Our analysis accounts for interaction effects and
the excitation of constants of motion which restrict the energy available
for thermalisation. 
\end{abstract}
\maketitle

\section{Introduction }

The harmonically trapped Bose-Einstein condensate (BEC) system is
well-isolated from its environment and the thermal state can be characterized by the parameters of total atom number ($N$), internal energy ($U$) and trap potential frequency ($\omega$). To
date the internal energy of a Bose gas has been impractical to measure experimentally with a useful
degree of accuracy. On the other hand, when a discernible thermal
fraction is present the temperature can be quite accurately  determined
by absorption imaging after expansion \cite{Ketterle1999a}. For this
reason the temperature dependence of BEC thermodynamics, e.g. condensate
fraction versus temperature, is quite well known, whereas the energy
dependence has hardly been studied. The few experimental studies conducted
have suffered from large uncertainties such that any form of quantitative
comparison with theory was not possible \cite{Ensher1996a}. More
recent studies have used an increase in temperature to signify the
imparting of energy \cite{Schori2004a,Kohl2005a},
but have not considered the quantitative relationship between these
two quantities.

Knowledge of the energy dependence of BEC thermodynamics is of wide
spread interest. The $\lambda$-transition in He was so named according
to the peculiar shape in the specific heat capacity of the system
\cite{Fairbank1958}.   Additionally a detailed description
of energy dependence would be useful for discriminating between
finite temperature theories of ultra-cold Bose gases.
For the case of degenerate Fermi gases a heat capacity measurement has been made by the Duke group \cite{Kinast2005a} by manipulating the trap potential in a manner similar to what we consider here for Bose gases. Additionally, recent work by the Heidelberg group \cite{Gati2006} has examined a precise method for measuring the temperature, and used this to confirm the deviation of the heat capacity of a Bose gas from that of a classical gas for a constant background noise source. However, their input heating rate was unknown, so that they were unable to quantify the heat capacity. The MIT group \cite{Onofrio2000b,Raman2001a} have measured  heating in a BEC by stirring it with a blue-duned focused light beam, and used those measurements  to distinguish between theories for drag forces.  
 
Here we propose using two mechanisms for transferring a precise amount
of energy to a BEC at near zero temperature to establish the relationship
between energy and temperature. For making calorimetric measurements,
one ideally would like a well-defined reservoir to transfer heat to the system
of interest. The isolation of ultra-cold atom experiments make such
an approach impractical, however an irreversible work process, such
as that done by a spinning paddle wheel in a fluid, is a convenient
method for transferring energy into these systems without changing
the external constraints.

Our main concern in this proposal is to develop and analyse  precise
ways of imparting energy to the system, and to characterize the portion
of this energy that is irreversible. While our analysis here focusses
upon Bragg scattering, and expansion from a trap, one could envisage doing this with other
methods, e.g. general perturbations of the trapping potential or stirring with
a focused light field. The main requirement is that the energy transfer is accurately calculable.

\section{Precise energy transfer}

\subsection{Overview of proposal}

In the next subsections we analyze two methods for precisely transferring energy ($\mathcal{E}_{\rm{trans}}$) to the system to be rethermalized. In Sec. \ref{BSsec} we consider the use of Bragg scattering, and in  Sec. \ref{EXPsec} we consider the sudden expansion from a harmonic trap. 
Having added this energy, the final temperature of the system (after it has returned to equilibrium) can be accurately measured. Knowledge of the irreversible work done on
the system and the rethermalized temperature establishes an equation
of state relationship of the form $T(U)$, where $T$ is the temperature
and $U$ is the internal energy (relative to the energy of the $T=0$ ground state). 
An important consideration in equating $U$ to $\mathcal{E}_{\rm{trans}}$, is that $\mathcal{E}_{\rm{trans}}$ must only consist of  the irreversible work done on the system. In particular, energy transferred to the Kohn mode (of harmonically trapped gases) needs to be excluded (as we discuss below). 

\subsection{Zero temperature formalism}
We consider our initial system to  be a Bose gas at zero temperature, where the condensate
is essentially pure. The condensate orbital satisfies the time-independent
Gross-Pitaevskii equation
\begin{equation}
\mu\Psi_{g}=\hat{H}_{{\rm sp}}\Psi_{g}+NU_{0}|\Psi_{g}|^{2}\Psi_{g},\label{tiGPE}
\end{equation}
 where $\hat{H}_{{\rm sp}}=\hat{p}^{2}/2m+V_{{\rm H}}(\mathbf{x})$
is the single particle Hamiltonian with 
\begin{equation}
V_{{\rm H}}(\mathbf{x})=\frac{1}{2}m\sum_{j=1}^3\omega_j^2x_j^2,
\end{equation}
the harmonic trapping potential, and $\{\omega_1,\omega_2,\omega_3\}$ are the trap frequencies along the coordinate directions. The quantity $\mu$ is the chemical
potential, $N$ the number of particles in the initial (pure) condensate, and $U_{0}=4\pi a\hbar^{2}/m$
is the interaction strength, where $a$ is the s-wave scattering length.

The energy of the ground state is given by the energy functional\begin{equation}
\mathcal{E}{}_{g}=E[\Psi_{g}]=\int d^{3}\mathbf{x}\, N\Psi_{g}^{*}\left[\hat{H}_{{\rm sp}}+\frac{NU_{0}}{2}|\Psi_{g}|^{2}\right]\Psi_{g}.\label{GPEfn}\end{equation}
 Many-body corrections to the ground state will in general be important,
however our interest here lies in understanding how much energy is
transferred to the system rather than the absolute energy, for which
the energy functional will suffice.

On several occasions we will have cause to make use of the 
Thomas-Fermi approximate solution to the Gross-Pitaevskii equation (\ref{tiGPE})  
\begin{equation}
|\Psi_g|^2 \approx|\Psi_{\rm{TF}}(\mathbf{x})|^2= \left\{
\begin{array}{cc} \frac{\mu_{\rm{TF}} - V_{\rm{H}}(\mathbf{x}) }{NU_0}, &V_{\rm{H}}<\mu_{\rm{TF}} ,\\
0, & \rm{elsewhere}
\end{array}\right\},\label{TFGS}
\end{equation}
with the chemical potential determined by
\begin{equation}
\mu_{\rm{TF}} = \frac{\hbar\bar{\omega}}{2}\left(15 N a 
\sqrt{\frac{m \bar{\omega}}{\hbar}}\right)^{2/5},
\end{equation}
where  $\bar{\omega} = (\omega_1 \omega_2\omega_3)^{1/3}$ (e.g. see \cite{Dalfovo1999a}).
This approximation is found by neglecting the kinetic term in the Gross-Pitaevskii equation, which is usually much smaller than the potential and interaction contributions. Comparisons with experiments have shown the Thomas-Fermi approximation to be a good description of  $T=0$ condensates (improving in accuracy as the number of particles in the condensate increases, e.g. see \cite{Dalfovo1999a}).

The Thomas-Fermi ground state has an energy of 
\begin{equation}
\mathcal{E}_g=\frac{5}{7}N\mu_{\rm{TF}},\label{ETF}
\end{equation}
calculated from Eq. (\ref{GPEfn}), with the kinetic term neglected.

\subsection{Method I: Energy Transfer by Bragg scattering}\label{BSsec}
Our first scheme for imparting energy uses the well-understood process of Bragg scattering \cite{Kozuma1999a,Stenger1999a,Blakie2001a} that is routinely used in labs for manipulating BECs.
We consider the situation where a Bragg pulse is used to first-order Bragg scatter a fraction $\alpha$ of the condensate (at rest) to momentum state
$\hbar\mathbf{b}$ (where $\mathbf{b}$ is the reciprocal lattice
vector of the Bragg potential). Note that the harmonic trap remains on during the Bragg scattering and the subsequent dynamics of the system as it rethermalizes. We also assume that the duration and intensity
of the Bragg potential are chosen so that all other orders of scattering
can be neglected, yet the scattering can be considered approximately
instantaneous on the timescale of condensate evolution (e.g. see Ref. \cite{Blakie2000a}). The
matter wave field at the conclusion of the scattering is then given
by\begin{equation}
\Psi_{i}(\mathbf{x})=\sqrt{1-\alpha}\Psi_{g}(\mathbf{x})+\sqrt{\alpha}e^{i\mathbf{b}\cdot\mathbf{x}}\Psi_{g}(\mathbf{x}),\label{eq:Psi_i}\end{equation}
 where we have assumed that the size of the condensate is large compared
to $1/\mathbf{b}$, so that the wave-packets centered at momenta $\mathbf{0}$
and $\hbar\mathbf{b}$ are orthogonal.

\subsubsection{Kohn Mode}

The initial state (\ref{eq:Psi_i}) has a momentum expectation of
$\mathbf{p}_{i}=\alpha\hbar\mathbf{b}$ per particle and according to Kohn's theorem
\cite{KohnThrm} this will lead to an undamped dipole oscillation
in the harmonic external potential. Transforming to the (non-inertial)
time-dependent centre-of-mass frame of reference, this oscillation
can be removed, while leaving the Hamiltonian for the system unchanged
(see \cite{KohnThrm}). In making this transformation an amount of
energy corresponding to the energy of the dipole oscillation, i.e.
\begin{equation}
E_{D}=Np_{i}^{2}/2m=\alpha^{2}N\hbar\omega_{b},\end{equation}
 is removed, where $\omega_{b}=\hbar b^{2}/2m$ is the Bragg recoil
frequency. Because this energy is locked into the centre of mass oscillation
it is not available for rethermalization.

However, if the motion of the scattered atoms enters into a sufficiently
anharmonic region of the trapping potential, then the Kohn mode energy
will be available to rethermalize. We expect this to be strongly dependent
on the manner in which the harmonic potential is made, but this effect
should be clearly observable as a decay in the centre-of-mass (COM) oscillation of
the system.

\subsubsection{Transferred energy}

The energy transferred to the condensate in the lab frame by the Bragg
scattering is calculated in Appendix \ref{sec:Interaction-energy}.
After subtracting the energy locked into the Kohn mode we obtain that
transferred energy available for rethermalization is \begin{equation}
\mathcal{E}_{{\rm trans}}\simeq N\hbar\omega_{b}(\alpha-\alpha^{2})+2\mathcal{E}_{{\rm int}}^{0}(\alpha-\alpha^{2}),\label{eq:transfEnergy}\end{equation}
 where   $\mathcal{E}_{{\rm int}}^{0}=(N^{2}U_{0}/2)\int d^{3}\mathbf{x}\,|\Psi_{g}(\mathbf{x})|^{4}$
is the interaction energy of the ground state. In the term proportional
to $\hbar\omega_{b}$ in Eq. (\ref{eq:transfEnergy}), the $\alpha$
contribution arises from the transfer of kinetic energy to the particles,
where as the $-\alpha^{2}$ part accounts for the energy locked into
the Kohn mode. The term proportional to $2\mathcal{E}_{{\rm int}}^{0}$
describes the additional energy arising from interactions due to the
creation of a coherent density fluctuation in the system. We have
taken this term to lowest order in the small parameter $\lambda/L$,
where $\lambda=2\pi/b$ is the wavelength of the density fluctuation
and $L$ is the size of the condensate. For more details we refer
the reader to Appendix \ref{sec:Interaction-energy}. The size of
these kinetic and interaction contributions to the energy are compared
in Table \ref{cap:Typical Values}. We note that for both cases considered
the interaction contribution is $\lesssim7\%$ the kinetic contribution.
So for many cases ignoring the interaction term in Eq. (\ref{eq:transfEnergy}) will be a good first approximation.

We note that the maximum transfer of energy occurs when $\alpha=1/2$,
since the energy available for rethermalization is proportional to
the momentum spread in the initial state (\ref{eq:Psi_i}) which is
maximized for $50\%$ scattering. For $\alpha>1/2$
the energy transferred by the Bragg scattering is increasingly locked
into the Kohn mode.

\begin{table}
\begin{tabular}{|c||c|c|}
\hline 
Atom&
$\hbar\omega_{b}$&
$2\mathcal{E}_{{\rm int}}^{0}/N$\tabularnewline
\hline
\hline 
$^{23}$Na &
$6.6\times10^{-29}$J$\approx 4.8\mu$K&
$4.3\times10^{-31}$J$\approx 0.031\mu$K\tabularnewline
\hline 
$^{87}$Rb&
$9.8\times10^{-30}$J$\approx 0.71\mu$K&
$7.3\times10^{-31}$J$\approx 0.053\mu$K\tabularnewline
\hline
\end{tabular}

\caption{\label{cap:Typical Values} Energy scales per particle for typical experimental
parameters. 
Each case considers a $10^6$ atom condensate in an isotropic $50$ Hz harmonic trap. We have taken the usual scattering lengths
for each atom \cite{Dalfovo1999a} and have taken $b$ to be that
for counter-propagating light fields of wavelength $\lambda=589$
nm and $\lambda=789$nm for the sodium and rubidium cases respectively
(i.e. $b=4\pi/\lambda$). }
\end{table}

\subsection{Method II: Energy Transfer by Expansion from trap}\label{EXPsec}
Our second method of energy transfer is to suddenly turn off the harmonic trapping potential (of initial frequencies $\{\omega_j\}$) for a period of time $t_{\rm{on}}$, allowing the condensate to expand, before the potential is reinstated (with final frequencies $\{\omega^\prime_j\}$).

To quantify the energy transfer in this process we need to consider the condensate dynamics after trap release. Castin and Dum have shown that such a system will undergo a \emph{self-similar expansion} \cite{Castin1996a}.   In particular,  if the trapping frequencies are time dependent, then
the condensate density remains as a Thomas-Fermi profile, but with 
time-dependent widths that evolve according to
\begin{eqnarray}
R_j(t) = \lambda_j(t) R_j(0),
\end{eqnarray}
where the equation of motion for the $\lambda_j$ are
\begin{eqnarray}
\ddot{\lambda}_j = \frac{\omega_j^2(0)}{\lambda_j \lambda_1 \lambda_2 \lambda_3}
 - \omega_j^2(t){\lambda}_j,\label{EOMlambda} 
\end{eqnarray}
with ${\lambda}_j(t=0) = 1$.

\subsubsection{Kohn Mode}
When the trap is turned off at $t=0$ the condensate begins to expand and it will also fall a distance $d = a_g t^2/2$ and acquire a velocity $v=a_gt$, where $a_g$ is the acceleration due to gravity.  
At time $t_{\rm{on}}$, when the trap is restored, $d$ and $v$ will manifest themselves as energy locked into the dipole mode\footnote{We note that the pure expansion is symmetric and does not couple to the dipole mode. Any energy locked into the dipole mode will be due solely to the effect gravity and changes in the trap equilibrium position.}.  We note that, depending on how the harmonic trap is produced, the trap centre may also change if the trap frequencies of the final trap are different to those of the initial trap. For this reason  we do not give explicit expressions for the dipole energy here. 

\subsubsection{Transferred energy}\label{unbndenergy}
Using the results of the  Appendix \ref{sec:Interaction-energy2}, we find that the energy transferred and available for thermalisation is 
\begin{equation}
\mathcal{E}_{\rm{trans}}=\frac{N\mu_{\rm{TF}}}{7}\left(2-5\bar{\gamma}^{6/5}+\sum_{j=1}^3\gamma_j^2\lambda_j^2(t_{\rm{on}})\right),\label{ETrapsTrapGen}
\end{equation}
where $\gamma_j=\omega_j^\prime/\omega_j$, $\bar\gamma=\sqrt[3]{\gamma_1\gamma_2\gamma_3}$ and $\mu_{\rm{TF}}$ is the Thomas-Fermi chemical potential of the initial condensate. Unlike the Bragg case, the transferred energy is unbounded since the condensate can be allowed to expand for arbitrarily long time periods.

\subsubsection{Analytic solution}\label{SecAnalytic}
While the energy transfer generally requires us to solve the ordinary differential equations (\ref{EOMlambda}), an approximate solution exists for the case of an elongated (cigar) trap. Here
the parameters $\lambda_\perp$ and $\lambda_z$ specify the system at time $t$.
Defining $\epsilon = \omega_\perp / \omega_z$ and $\tau = \omega_{\perp} t$
we can find an approximate
  solution \cite{Castin1996a} 
\begin{eqnarray}
\lambda_{\perp} &=& \sqrt{1 + \tau^2},\\
\lambda_{z} &=& 1 + \epsilon^2[\tau \tan^{-1} \tau -\ln \sqrt{1 + \tau^2}]
+ O(\epsilon^4).
\end{eqnarray}
Thus the energy available for thermalisation given the same initial and final trap parameters is
\begin{eqnarray}
\mathcal{E}_{\rm{trans}}& = & 
\frac{2 N\mu_{\rm{TF}}}{7}\left(\tau^2_{\rm{on}} +
\epsilon^2[\tau_{\rm{on}} \tan^{-1} \tau_{\rm{on}}  \right.\nonumber\\&&\left.-\ln \sqrt{1 + \tau^2_{\rm{on}}}] + O(\epsilon^4)
\right),\label{ETrapsTrapAnalytic}
\end{eqnarray}
 with $\tau_{\rm{on}}=\omega_{\perp} t_{\rm{on}}$.

\section{Application of calorimetry}
 We now consider the application of our calorimetry scheme to an ideal trapped Bose gas with critical temperature given by $T_{c}=\frac{\hbar\bar{\omega}}{k_{B}}\left[{N}/{\zeta(3)}\right]^{1/3}$.
For $T<T_{c}$,
we have that the energy of this gas is given by
\begin{equation}
U(T)=3Nk_{B}\frac{\zeta(4)}{\zeta(3)}\frac{T^{4}}{T_{c}^{3}},\end{equation}
 where $\zeta(\alpha)=\sum_{n=1}^{\infty}n^{-\alpha}$ (e.g. see
\cite{Giorgini1996a}).
\subsection{Bragg limitations}
In order for the Bragg scheme to be capable of probing up to the transition
region, we require that maximum energy transfer is greater than the
energy content of the gas at the critical point (i.e. $\mathcal{E}_{{\rm trans}}(\alpha=1/2)>U(T_{c})$).
Evaluating this inequality sets the following constraint on the system
size and trap
\begin{equation}
\hbar\bar{\omega}\sqrt[3]{N}<\frac{\hbar\omega_b}{12}\frac{\zeta(3)^{4/3}}{\zeta(4)}.
\end{equation}
 Thus for $N$ or $\bar\omega$ too large, first order Bragg \textbf{}will
not provide sufficient energy to heat the BEC from $T=0$ to $T=T_{c}$.
For example, $150\times10^{3}$  rubidium-87 atoms in a 25 Hz trap is about at this threshold.
For sodium-23, the higher kinetic energy of the Bragg scattering ($\hbar\omega_b$)
allows larger/tighter systems to be used. One could circumvent this
limitation by using second order Bragg scattering to impart a larger
amount of energy.

We note that given the high recoil temperatures (relative to usual $T_{c}$ values)
associated with the light used to Bragg scatter it may seem surprising
that it is not always possible to Bragg scatter sufficient energy
to take the system to the critical point. However, it should be kept
in mind that the recoil temperature is defined for an atom with a mean
energy equal to the recoil energy in each degree of freedom, whereas for
our Bragg scattering process this energy must be shared between the 6 degrees
of freedom of the atom (i.e. 3 kinetic and 3 potential), and at most only 50\% of the total energy transferred
is available for rethermalization.

\subsection{Error analysis of Bragg energy transfer}

\begin{figure}
\includegraphics[width=8.5cm]{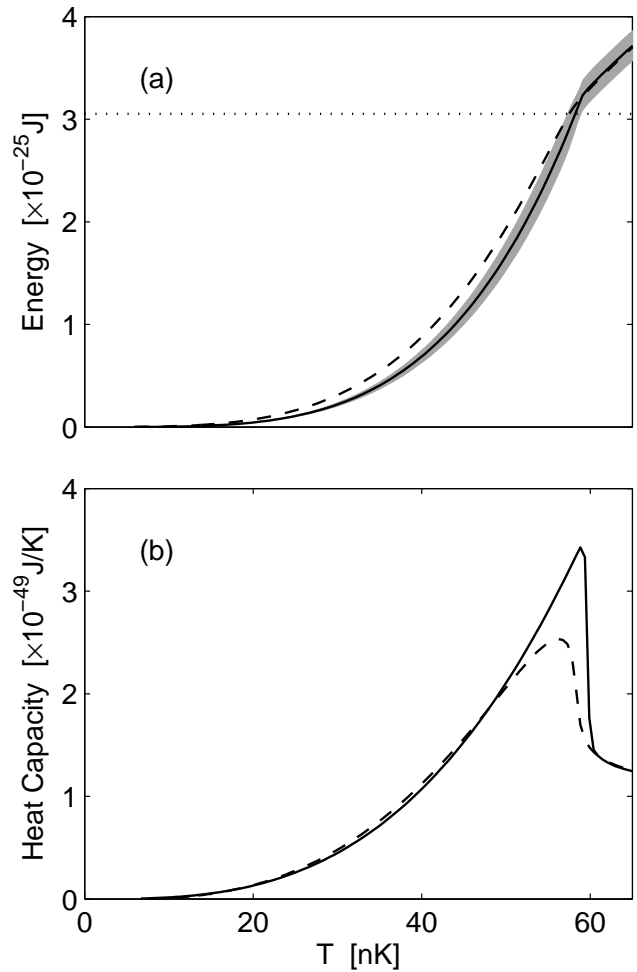} 
\caption{\label{cap:Fig1} (a) Internal energy ($U$) versus temperature. 
(b) The specific heat capacity ($\Delta U/\Delta T$). Ideal gas (solid line), Hartree-Fock Bogoliubov-Popov calculation of interacting gas (dashed).
Energy corresponding to $\alpha=0.5$ is shown as horizontal dotted line in (a) and grey region  indicates the uncertainty in the transferred
energy for the ideal case given that there is a 5\% uncertainty in the total number of
atoms (i.e. $\Delta N=0.05\times N$), the fraction scattered is
accurate to the 5\% level (i.e. $\Delta\alpha=0.05\times\alpha$).
System parameters: $1.5\times10^{5}$ Rb atoms in an isotropic 25 Hz trap. }
\end{figure}

Error in the transferred energy will likely be dominated by shot-to-shot
variation in the number of atoms in the initial condensate and error
in the fraction of scattered atoms. Here we denote these errors as
$\Delta N$ and $\Delta\alpha$ respectively and account for their
effect on our ability to know the precise amount of energy transferred
to the system. Linearizing Eq. (\ref{eq:transfEnergy}), we find that
the energy transferred is affected by these quantities according to\begin{eqnarray}
\Delta\mathcal{E}_{{\rm trans}} & = & \hbar\omega_{b}\left[(\alpha-\alpha^{2})\Delta N
 +N(1-2\alpha)\Delta\alpha\right],\end{eqnarray}
 where we have neglected the contributions of the interaction term since
it is typically an order of magnitude smaller. In Fig. \ref{cap:Fig1}
we consider the effect of these errors in the use of our scheme to
determine the heat capacity.

Additional error will arise because the initial condensate temperature
is non-zero, however since the energy is a rapidly increasing function
of $T$ %
\footnote{For the ideal Bose gas  $U\sim T^{4}$.}, this error is typically several orders of magnitude smaller than
the errors due to uncertainty in atom number and scattered fraction.

In Fig. \ref{cap:Fig1} we also compare $U(T)$ for the interacting versus ideal system. The interacting properties are calculated using the Hartree-Fock-Bogoliubov theory in the Popov approximation (e.g. see \cite{Giorgini1997,Minguzzi1997,Gerbier2004b,Hutchinson1997a,Davis2006a}). We see that for $T<T_c\sim60$ nK the dependence of $U$ on $T$ for the interacting system  is noticeably distinguishable from the ideal gas (i.e. beyond the limits of the error in energy transfer). This suggests that interaction effects could be experimentally measured using this technique if  sufficiently good reproducibility of initial condensate number and Bragg scattering precision can be obtained.

\subsection{Error analysis of energy transfer by trap expansion}
As discussed in section \ref{unbndenergy} imparting energy by trap expansion has no upper bound to the amount of transferred energy.  We would expect that the shot to shot variation in atom numbers will dominate the error budget, however there may be other important considerations relating to the particular way the trap is produced.

For the analytic case given in Sec. \ref{SecAnalytic}, we examine the sensitivity of the energy transfered to variations in atom number ($\Delta N$) and errors in the axial ($\Delta\omega_z$) and radial ($\Delta\omega_\perp$) trapping frequencies. By linearizing  Eq. (\ref{ETrapsTrapAnalytic}) we find
\begin{eqnarray}
\Delta\mathcal{E}_{{\rm trans}}&=&\frac{7}{5}\mathcal{E}_{\rm{trans}}\frac{\Delta N}{N} +\left(\frac{4}{7}N\mu_{\rm{TF}}\tau^2_{\rm{on}}-\frac{8}{5}\mathcal{E}_{\rm{trans}}\right)\frac{\Delta \omega_z}{\omega_z} \nonumber\\
&+&\left(\frac{2}{7}N\mu_{\rm{TF}}\epsilon^2\tau_{\rm{on}}\tan^{-1}\tau_{\rm{on}}+\frac{14}{5}\mathcal{E}_{\rm{trans}}\right)\frac{\Delta \omega_{\perp}}{\omega_{\perp}}.\nonumber\\
&&\label{EtrapUncert}
\end{eqnarray}
Assuming that the dominant uncertainty is atom number variation, we see that the relative uncertainty in transferred energy is roughly proportional to the relative uncertainty in atom number. For the general case given in Eq. (\ref{ETrapsTrapGen}), the error associated with trap uncertainties requires a numerical integration of the differential equation (\ref{EOMlambda}).
However, the (likely dominant ) error associated with number uncertainty is still  given by the first term in Eq. (\ref{EtrapUncert}).

\section{Conclusions}
We have presented two practical schemes for performing calorimetry on a Bose-Einstein condensate system.
It is clear from our results that reasonably accurate calorimetry
measurements could be made using Bragg scattering or by controlled expansion from a confining potential. We have characterized the sensitivity of these methods to typical experimental uncertainties in atom number and  have also shown that it
 should be feasible to measure  interaction effects on the thermal properties of a Bose gas.

Our scheme also presents a rather well-defined initial condition for studying non-equilibrium dynamics.
There is considerable interest in the dynamics of the thermalisation, as
we expect there will be a crossover from coherent to incoherent dynamics.
This topic is of significant current interest (e.g. see \cite{Cazalilla2006,Simula2006a,Kinoshita2006,Rigol2007,Kollath2007,Manmana2007,Cramer2007,Wuster2007a,Calabrese2007}) andwill be the subject of future work using classical field methods (e.g. \cite{Blakie2005,Blakie2007a}).

\appendix  
\section{Energy calculations}
\subsection{Bragg transferred energy\label{sec:Interaction-energy}}
Here we calculate the total energy transferred to the condensate by
Bragg scattering in the lab frame. This is found by evaluating $\mathcal{E}_{{\rm trans}}=E[\Psi_{i}]-E[\Psi_{g}]$.
Assuming that the original and scattered wave packets are well
separated in momentum space, we can make the approximations \begin{eqnarray}
\int d^{3}\mathbf{x}\,|\Psi_{g}|^{2}e^{i\mathbf{b}\cdot\mathbf{x}} & \approx & 0,\\
\int d^{3}\mathbf{x}\,|\Psi_{g}|^{4}e^{i\mathbf{b}\cdot\mathbf{x}} & \approx & 0,\\
\int d^{3}\mathbf{x}\,|\Psi_{g}|^{2}e^{2i\mathbf{b}\cdot\mathbf{x}} & \approx & 0.\end{eqnarray}
 These integrals are all of order $\lambda/L$ where $L$ is the spatial
size of the condensate and $\lambda=2\pi/b$ is the wavelength of
the Bragg induced density modulation. In experiments typical values
are $\lambda/L\sim1/50$, so these approximations are well satisfied
and higher order terms can be ignored.

With the above approximations we obtain \begin{equation}
E[\Psi_{i}]=\mathcal{E}_{{\rm sp}}^{0}+\alpha N\hbar\omega_{b}+(1+2\alpha-2\alpha^{2})\mathcal{E}_{{\rm int}}^{0},\end{equation}
 where \begin{eqnarray}
\mathcal{E}_{{\rm sp}}^{0} & = & N\int d^{3}\mathbf{x}\,\Psi_{g}^{*}\hat{H}_{{\rm sp}}\Psi_{g},\\
\mathcal{E}_{{\rm int}}^{0} & = & \frac{N^{2}U_{0}}{2}\int d^{3}\mathbf{x}\,|\Psi_{g}|^{4},\label{E0int}\end{eqnarray}
 are the single-particle and interaction energies of the ground state
respectively, with $\mathcal{E}_{g}=\mathcal{E}_{{\rm sp}}^{0}+\mathcal{E}_{{\rm int}}^{0}$.
Thus the transferred energy is given by \begin{equation}
\mathcal{E}_{{\rm Bragg}}=\alpha N\hbar\omega_{b}+2(\alpha-\alpha^{2})\mathcal{E}_{{\rm int}}^{0}.\end{equation}

The Thomas-Fermi approximation to the ground state (\ref{TFGS}) gives
  \begin{equation}
\mathcal{E}_{{\rm int}}^{0}=\frac{2}{7}N\mu_{\rm{TF}}.\end{equation}

\subsection{Trap expansion transferred energy\label{sec:Interaction-energy2}}
Having solved Eqs. (\ref{EOMlambda}), the condensate density is then given by
\begin{eqnarray}
|\Psi(\mathbf{x},t)|^2 = \frac{\mu_{\rm{TF}} - \sum_j \frac{1}{2} m 
 \omega_j(0)^2 x_j^2/\lambda_j^2(t)}{NU_0 \lambda_1(t) \lambda_2(t) \lambda_3(t)}.
\label{eqn:td_density}
\end{eqnarray} 

\subsubsection{Expansion}

We envisage turning the trap off suddenly.  In the Thomas-Fermi approximation,
for the ground state, the kinetic energy is negligible meaning that the system only has interaction
energy given by the generalized form of Eq. (\ref{E0int})
\begin{equation}
\mathcal{E}_{\rm{int}}(t) = \frac{N^2U_0}{2} \int d^3\mathbf{x} |\Psi(\mathbf{x},t)|^4.
\end{equation}
The condensate begins to expand: the interaction energy is converted to kinetic
energy $(\mathcal{E}_{\rm{kin}})$.  The energy balance is determined by $\mathcal{E}_{\rm{kin}}(t) =  \mathcal{E}_{\rm{int}}(0) - \mathcal{E}_{\rm{int}}(t)$, and using the time-dependent density of Eq.~(\ref{eqn:td_density}) we find
\begin{equation}
\mathcal{E}_{\rm{int}}(t) = \frac{2N \mu_{\rm{TF}} }{7 \lambda_1(t) \lambda_2(t) \lambda_3(t)}.
\end{equation}

Now consider turning the trap back on suddenly at time $t=t_{\rm{on}}$, with some possibly different trap frequencies $\{\omega_j'\}$.   Defining the ratio of new to old  trapping frequencies as $\gamma_j = \omega_j' / \omega_j$,  the potential energy   at the instant the new potential is turned on  is given by
\begin{eqnarray}
\mathcal{E}_{\rm{pot}}' &=&\int d^3\mathbf{x} V'_{\rm{H}}(\mathbf{x}) |\Psi(\mathbf{x},t_{\rm{on}})|^2,
\\
&=&\frac{N\mu_{\rm{TF}} }{7}\sum_{j=1}^3\gamma_j^2 \lambda_j^2(t_{\rm{on}}).
\end{eqnarray}
and the total energy of the system will be the sum of the interaction energy at $t=0$ and the potential energy at $t=t_{on}$, i.e.
\begin{eqnarray}
\mathcal{E}_{\rm{tot}}' = \frac{N\mu_{\rm{TF}} }{7}\left(2 + \sum_{j=1}^3\gamma_j^2 \lambda_j^2(t_{\rm{on}})
\right).
\end{eqnarray}
The  using the Thomas Fermi result in Eq. (\ref{ETF}) we find the approximate ground state energy in the final trapping potential  
\begin{equation}
\mathcal{E}_{g}'  = \frac{5 N\mu_{\rm{TF}} \bar{\gamma}^{6/5} }{7}
\end{equation}
where we have defined $\bar{\gamma}$ as the geometric mean of
the $\gamma_j$.  Thus the energy added by the sudden release and re-application of the harmonic trapping potential is
\begin{eqnarray}
\mathcal{E}_{\rm{exp}} & = & \mathcal{E}_{\rm{tot}}' - \mathcal{E}_{g}', \\
 & = & \frac{N\mu_{\rm{TF}}}{7}\left(2 
 - 5\bar{\gamma}^{6/5}+ \sum_{j=1}^3\gamma_j^2 \lambda_j^2(t_{\rm{on}})
\right).
\end{eqnarray}

\bibliographystyle{unsrt.bst}   
\bibliography{Ethermo}

\end{document}